\documentclass[sigconf,authorversion]{acmart}

\AtBeginDocument{%
  }

\copyrightyear{2025}
\acmYear{2025}
\setcopyright{none}
\acmConference[CHI EA '25]{Extended Abstracts of the CHI Conference on
Human Factors in Computing Systems}{April 26-May 1, 2025}{Yokohama, Japan}
\acmBooktitle{Extended Abstracts of the CHI Conference on Human Factors in
Computing Systems (CHI EA '25), April 26-May 1, 2025, Yokohama,
Japan}
\acmDOI{10.1145/3706599.3719798}
\begin{document}

\title{VizTrust: A Visual Analytics Tool for Capturing User Trust Dynamics in Human-AI Communication}

\author{Xin Wang}
\authornote{Corresponding author.}
\affiliation{%
  \institution{School of Systems Science and Industrial Engineering, Binghamton University}
  \city{Vestal}
  \state{New York}
  \country{USA}}
\email{xwang314@binghamton.edu}
\orcid{0009-0006-7395-6988}

\author{Stephanie Tulk Jesso}
\affiliation{%
  \institution{School of Systems Science and Industrial Engineering, Binghamton University}
  \city{Vestal}
  \state{New York}
  \country{USA}
}
\email{stulkjesso@binghamton.edu}
\orcid{0000-0003-0605-4782}

\author{Sadamori Kojaku}
\affiliation{%
  \institution{School of Systems Science and Industrial Engineering, Binghamton University}
  \city{Vestal}
  \state{New York}
  \country{USA}}
\email{skojaku@binghamton.edu}
\orcid{0000-0002-9414-6814}

\author{David M Neyens}
\affiliation{%
  \institution{Department of Industrial Engineering, Clemson University}
  \city{Clemson}
  \state{South Carolina}
  \country{USA}}
\email{dneyens@clemson.edu}
\orcid{0000-0002-3443-518X}

\author{Min Sun Kim}
\affiliation{%
 \institution{School of Communication and Information, University of Hawaii at Manoa}
 \city{Honolulu}
 \state{Hawaii}
 \country{USA}}
\email{kmin@hawaii.edu}
\orcid{0000-0002-0027-1383}

\renewcommand{\shortauthors}{Wang et al.}

\begin{abstract}
Trust plays a fundamental role in shaping the willingness of users to engage and collaborate with artificial intelligence (AI) systems. Yet, measuring user trust remains challenging due to its complex and dynamic nature. While traditional survey methods provide trust levels for long conversations, they fail to capture its dynamic evolution during ongoing interactions. Here, we present VizTrust\footnote{VizTrust will be publicly accessible, which can be tracked at \href{https://github.com/xin-wang-kr/VizTrust-project}{github.com/xin-wang-kr/VizTrust-project}}, which addresses this challenge by introducing a real-time visual analytics tool that leverages a multi-agent collaboration system to capture and analyze user trust dynamics in human-agent communication. Built on established human-computer trust scales—competence, integrity, benevolence, and predictability—, VizTrust enables stakeholders to observe trust formation as it happens, identify patterns in trust development, and pinpoint specific interaction elements that influence trust. Our tool offers actionable insights into human-agent trust formation and evolution in real time through a dashboard, supporting the design of adaptive conversational agents that responds effectively to user trust signals.
\end{abstract}

\ccsdesc[500]{Human-centered computing~Natural language interfaces}
\ccsdesc[500]{Computer systems organization~Real-time systems}
\ccsdesc[500]{Human-centered computing~Visualization toolkits}

\keywords{Agentic System, Conversational Agents, Natural Language Processing, Visualization Analysis, Large Language Model, Machine Learning}
\begin{teaserfigure}
  \centering
  \includegraphics[width=0.94\textwidth]{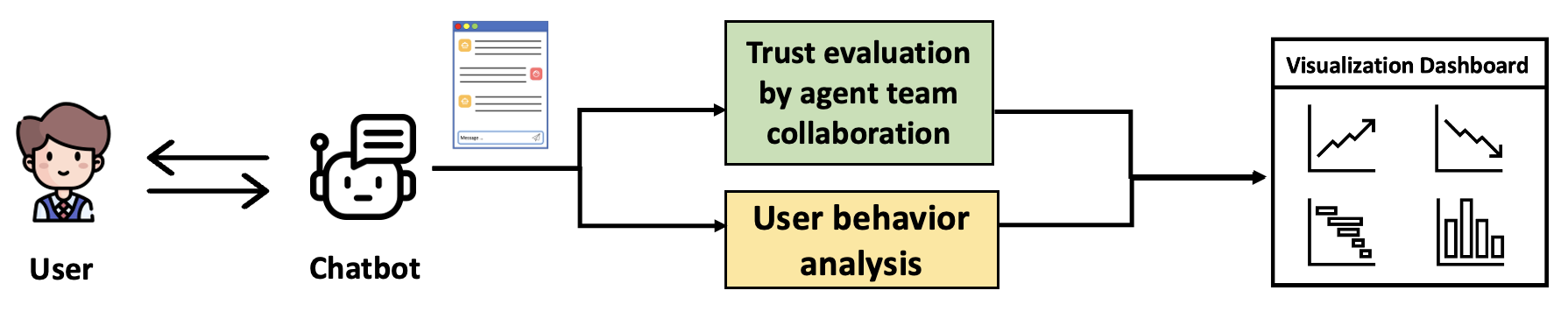}
  \caption{Workflow of VizTrust.}
  \Description{The VizTrust workflow shows that a user can interact with front-end chatbot and then the conversation transcripts will be passed down for trust evaluation by agent team collaboration and user behavior analysis in back-end system. The data collected on the back-end side will be visualized on dashboard.}
  \label{fig:teaser}
\end{teaserfigure}


\maketitle

\section{Introduction}
The increasing prevalence of conversational agents in diverse applications—from customer service to healthcare—has brought user trust to the forefront of human-agent communication research \cite{biro2023effects, seitz2022can, nordheim2019initial, harrington2023trust, cai2022impacts}. Understanding user trust is crucial for conversational agent design as it significantly affects effective interaction, shapes user satisfaction and engagement, and directs adoption of agent systems \cite{wald2021make, ruttkay2004brows, rheu2021systematic}. Yet, understanding how user trust develops and evolves during interactions remains notoriously challenging. User trust has multifaceted nature and can fluctuate over human-agent communication \cite{hoff2015trust, seymour2021exploring}. Traditional approaches, such as post-interaction surveys or qualitative analysis, fail to capture the fluid and temporal nature of trust development. 

This study proposes VizTrust, a visual analytics tool which can capture user trust dynamics in real-time human-agent communication process. VizTrust applies a multi-agent collaboration framework to decompose user trust evaluation task, based on four established trust dimensions, including competence, integrity, benevolence, and predictability \cite{mayer1995integrative, corritore2003line, haresamudram2024talking, bhattacharya1998formal, gulati2019design, muir1996trust, yamani2020human, pinto2022trust}. VizTrust also presents a variety of social signals in real-time analysis, including linguistic strategies, empathetic responses, and emotional tones, aiding conversational design stakeholders better observe human-agent communication process. 

\section{Related Work}
\subsection{User trust in human-AI communication}
User trust in AI-enabled systems has been studied extensively in human-computer interaction (HCI), social psychology, and organizational behavior \cite{bach2024systematic}. These studies highlight the multidimensional nature of trust, which can be influenced by factors related to the user, the AI system, and the broader socio-ethical context \cite{emaminejad2021trust}. From an HCI perspective, research has identified key trust-related characteristics such as transparency, reliability, and immediacy behaviors \cite{sousa2024human}. Moreover, the anthropomorphism of AI systems can influence emotional trust, while cognitive trust is associated with more tangible system features \cite{sousa2024human}. The literature also suggests that a human-centric approach is necessary to better understand the nuances of user trust, moving beyond narrow, technical solutions that fail to capture the complex and contextual nature of trust dynamics in human-AI interactions \cite{bach2024systematic, emaminejad2021trust}. Existing methods for measuring user trust in AI-enabled systems often rely on post-interaction surveys, interviews, and qualitative analyses; therefore, there is a significant opportunity to develop new methods that can yield insight into trust formation in real time.

\subsection{Connection between user trust perceptions and user behaviors }
The connection between user trust perceptions and user behaviors in AI interactions is a critical area of research, as trust influences various aspects of human-AI engagement \cite{bach2024systematic}. Trust affects how users interact with AI systems, impacting their engagement levels, communication styles, and willingness to follow recommendations  \cite{bach2024systematic, choudhury2023investigating}. Furthermore, trust is intertwined with politeness strategies and emotion, which, when appropriately leveraged, can enhance the quality of these interactions \cite{kumar2022politeness}. Politeness strategies, a well-established concept in sociolinguistics \cite{brown1987politeness}, are particularly relevant in human-AI interactions \cite{kumar2022politeness}. Polite linguistic strategies can mitigate social threats and signal respect for the other party in the interaction \cite{mansoor2018politeness}. When users perceive an AI system as polite, competent, and respectful, they are more likely to reciprocate these politeness behaviors \cite{bowman2024exploring}. Studies have demonstrated a strong link between user trust perceptions and behavioral responses during interactions with AI systems, including engagement levels, communication patterns, and adherence to AI recommendations \cite{sousa2024human, cascio2023drives, ajenaghughrure2018interplay}. Overall, the subtle nature of behavioral trust signals must necessitate further study.  

\section{VizTrust}
\subsection{Design goal}
Within this paper, we present VizTrust, a web-based application built using Streamlit framework. Our aim is to develop dynamic displays of trust, behavioral interactions, and engagement with AI systems to support real-time assessments and to identify the dynamic nature of assessment of user interactions with AI systems. In doing so, we support design stakeholders of conversational agents through the following design goals:

\textbf{G1. Real-time trust analysis:} Understanding moment-to-moment trust evolution is essential, as it provides immediate insights that static measurements cannot capture \cite{tang2014trust}. VizTrust should enable real-time trust assessment during human-agent communication to accurately portray overall user trust dynamics and identify trends of user trust, especially for longer human-agent conversations.

\textbf{G2. Multidimensional trust evaluation:} Multiple dimensions of trust provide a holistic understanding necessary for optimizing conversational agent design \cite{mayer1995integrative}. VizTrust should evaluate user trust across four key dimensions, including competence, integrity, benevolence, and predictability, incorporating social psychology principles. 

\textbf{G3. Key conversational metrics:} Subtle linguistic and behavioral signals reflect trust in real-world human-agent communication \cite{rheu2021systematic}. VizTrust should provide a comprehensive analysis of trust formation with the involution of key conversational metrics extracted from users' language interaction, such as user engagement metrics, politeness strategy usage, and emotional tones. 

\textbf{G4. Time series visualization analysis:} Critical trust turning points in interactions require targeted design improvements, particularly in long and complex conversations \cite{daronnat2020impact}. VizTrust should visualize the user trust dynamics and its associated metrics over the whole conversation between human user and agent, highlighting key moments where the conversational agent’s reaction triggers user trust changes in a long conversation.

\subsection{VizTrust workflow}
VizTrust contains a front-end interaction with human users and a back-end analysis system leading to comprehensive visualization analysis on human-agent communication. A front-end chatbot of VizTrust (set up with \texttt{Llama-3.1-8b} model) interacts with human users in natural languages on topics related to conversational agent design (Figure \ref{fig:teaser}). As receiving each user's utterance, the front-end chatbot passes the current conversation transcript to a multi-agent system that evaluates trust and Python programs for user behavior analysis using natural language processing and machine learning techniques. Both analysis processes are iteratively conducted by conversation turns separated by each user utterance that the chatbot receives, and then the outcomes over all conversation turns are collected and visualized on the dashboard for observations of design stakeholders. More details are explained in the following sections.

\subsection{VizTrust user interface}

\begin{figure*}[h]
  \centering
  \includegraphics[width=0.9\linewidth]{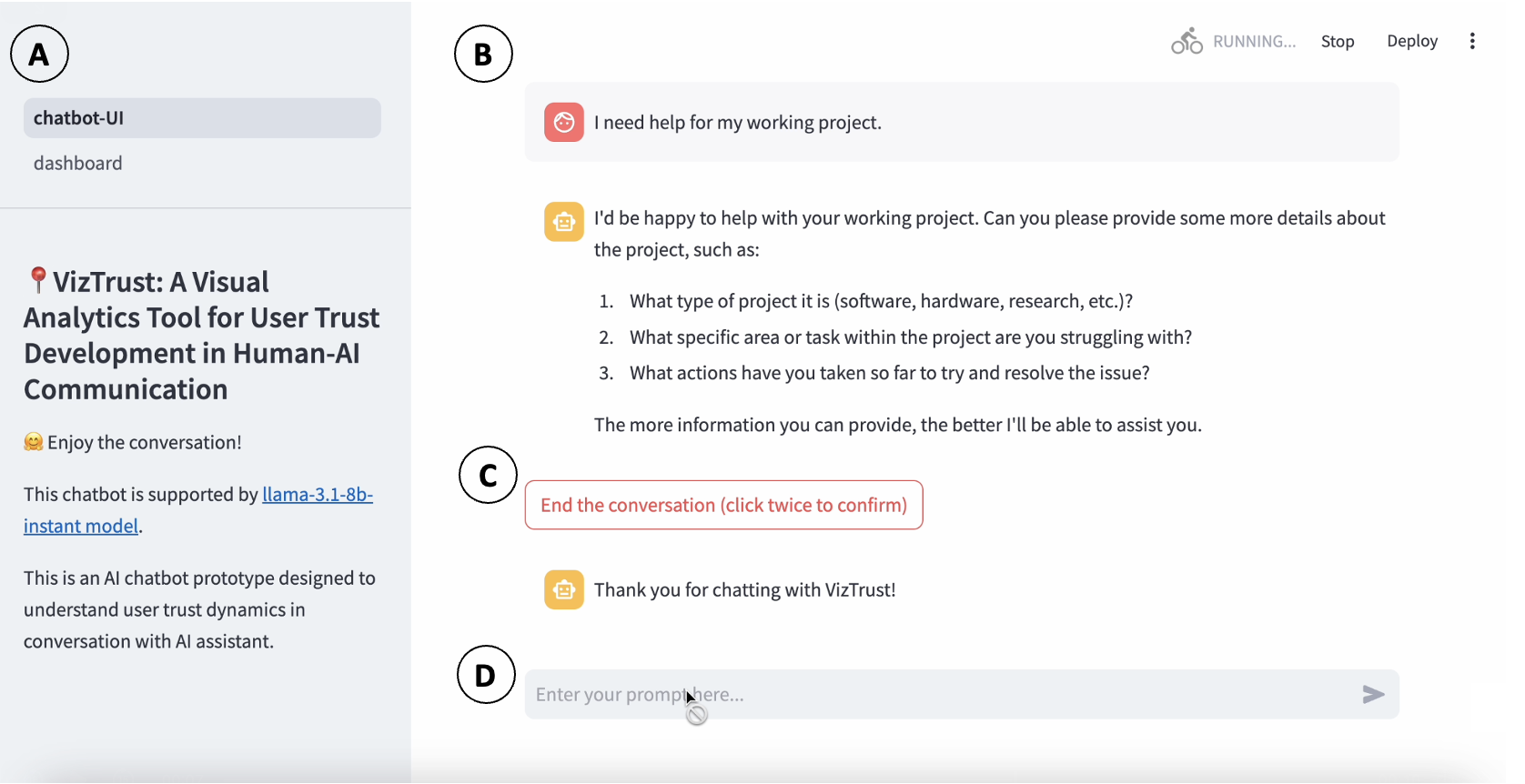}
  \caption{(A) The control panel enables the switch between chatbot user interface and visualization dashboard. (B) The chat window shows language interaction between user and chatbot. (C) "End the conversation" button allows the dashboard to display. (D) Prompt message window inputs user’s message text.}
  \Description{VizTrust features a chatbot user interface comprising a control panel, a chat window, a prompt message window, and an "End the Conversation" button. Testing users can input query messages to chatbot using the prompt message window and review the chatbot's responses in chat window. When testing users choose to conclude the conversation with chatbot, they press "End the conversation" button, after which the dashboard becomes accessible for analysis.}
  \label{fig:chatbot-ui}
\end{figure*}

VizTrust has a chatbot user interface for the invited human users and a visualization dashboard interface for the conversational agent design stakeholders. The initial page of VizTrust is set to the chatbot user interface for the first user test. These two user interfaces can be switched by control panel (Figure \ref{fig:chatbot-ui} (A)).

In chatbot user interface, users can input any query to chatbot through prompt message window (Figure \ref{fig:chatbot-ui} (D)) and get chatbot's textual response from chat window (Figure \ref{fig:chatbot-ui} (B)). Testing users can decide to stop the conversation at any moment by simply clicking the "End the conversation" button (Figure \ref{fig:chatbot-ui} (C)) twice to confirm. After users' ending confirmation, the chatbot will express appreciation for users' participation and the prompt message window will be locked, avoiding further prompt inputs. While the users also have access to the control panel, the dashboard will show a blank board before VizTrust receives ending confirmation. This is intended to prevent the situation where users read the analysis results and consciously give desired responses instead of being authentic.

\begin{figure*}[h]
  \centering
  \includegraphics[width=0.9\linewidth]{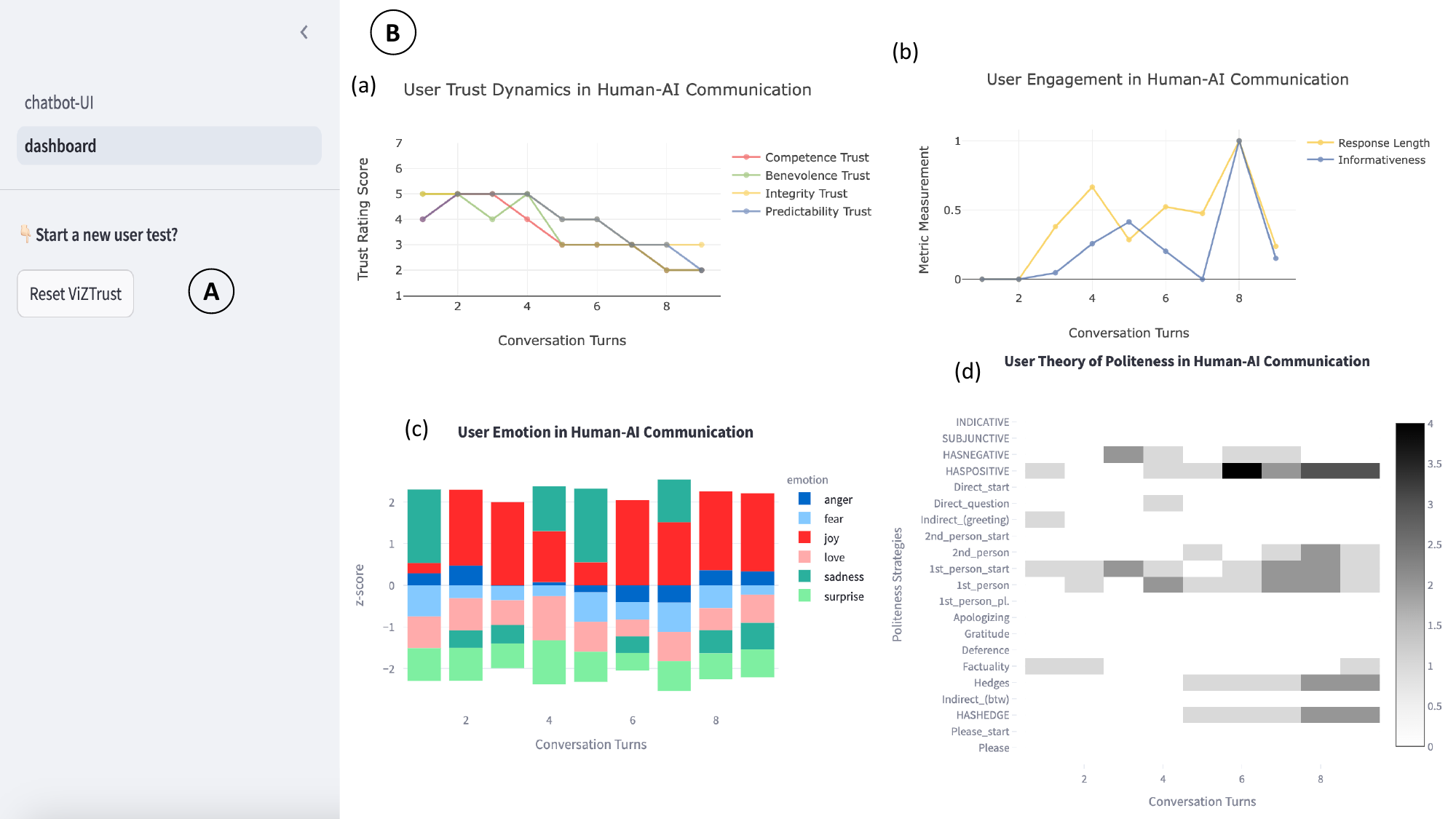}
  \caption{(A) “Reset VizTrust” button removes current user testing records to restore to initial state. (B) Dashboard shows interactive visualization.}
  \Description{VizTrust features a visualization dashboard interface that includes four time series visualizations and a "Reset VizTrust" button. Four time series visualizations present analysis across all conversation turns, covering user trust dynamics, user engagement, user emotional tones, and user theory of politeness. Design stakeholders can use the "Reset VizTrust" button to quickly erase the current user record and prepare for the next user test.}
  \label{fig:dashboard}
\end{figure*}

In the visualization dashboard interface, the main view (Figure \ref{fig:dashboard} (B)) presents four time series visualizations of all conversation turns, including user trust dynamics, user engagement, user emotional tones, and user theory of politeness. Each visualization plot has interactive features that allow design stakeholders to gain more detailed information on trend-changing points. The design stakeholders can select any conversation turn on the plot of user trust dynamics to read supporting evidence on the trust rating scores in detail (Figure \ref{fig:evaluation-summary} in Appendix). The “Reset VizTrust” button (Figure \ref{fig:dashboard} (A)) is designed for design stakeholders to quickly erase the current user record and prepare for the next user test.  

\subsection{Multi-agent system design for trust evaluation}


The multi-agent collaboration system enhances VizTrust's ability to comprehensively understand user trust in human-agent conversations through collective decision making \cite{baltaji2024conformity}. Leveraging insights from agents with diverse knowledge and expertise on specific trust dimensions can increase the adaptability of VizTrust to user trust complexity \cite{rosen2018teamwork, el2019shared}.

\subsubsection{LLM setting}

This system integrates the open-source LLM (large language model), \texttt{Mixtral-8x7B} (temperature = 0), offering cost-efficient and reliable performance for early-stage conversational agent design. \texttt{Mixtral-8x7B} outperforms closed-source models like GPT-3.5 Turbo and Claude-2.1 \cite{jiang2024mixtral, quan2024verification, nezhurina2024alice, auffarth2023generative}, making it well-suited for powering multi-agent collaboration system inside VizTrust application and facilitate efficient communication and decision-making performance among agents for complex tasks.

\subsubsection{Collaborative agent team design}
The multi-agent collaboration system consists of six conversable LLM agents (Figure \ref{fig:agent-team} in Appendix), built by AutoGen, an open-source programming framework for agentic AI \cite{wu2024autogen}. Each LLM agent is designed with a specific role in the agent team:

(1) Trust evaluation agent acts as a team leader. Its main role is to conversate with each of the four trust-specific agents in sequential turn to gather their feedback on conversation transcripts between a human user and the front-end chatbot of VizTrust. During these one-on-one interactions, the team leader asks trust-specific agent to rate the corresponding user trust dimension using a 7-point Likert scale (e.g., 1=extremely low; 2=low; 3=somewhat low; 4=neutral; 5=somewhat high; 6=high; 7=extremely high) and to provide supporting evidence. Based on responses from trust-specific agents, the team leader makes decisions to better understand the user’s trust during communication with chatbot.

(2) Social psychologist agent for competence trust is built with competence trust concept, which is the users’ belief in the AI system's ability, skills, and expertise to perform tasks effectively and accurately within its intended domain \cite{pinto2022trust}. This agent analyzes human users' language expression indicating their competence trust by checking if users positively accept advice, recognize expertise, and ask follow-up questions. 

(3) Social psychologist agent for integrity trust is built with integrity trust concept, which is the users’ belief in that the AI system adheres to a set of acceptable principles, is honest about its capabilities and limitations, and provides truthful and accurate information \cite{muir1996trust}. This agent analyzes human users’ language expression indicating their integrity trust by checking if users doubt the sources of chatbot's answer, request verification of information related to chatbot's answer, and show tolerance toward the chatbot’s limitation. 

(4) Social psychologist agent for benevolence trust is built with benevolence trust concept, which is the users’ belief in that the AI system acts in their best interest, shows genuine concern for their needs, and aims to provide helpful and beneficial assistance \cite{gulati2019design, pinto2022trust}. This agent analyzes human users’ language expression indicating their benevolence trust by checking if users self-disclose verbally, give emotional sharing, and proactively seek guidance. 

(5) Social psychologist agent for predictability trust is built with predictability trust concept, which is the users’ belief in that the AI system's behaviors and responses are consistent, follow understandable patterns, and meet expected standards across interactions \cite{pinto2022trust}. This agent analyzes human users’ language expression indicating their predictability trust by checking if users mentioned the past satisfied interactions.

(6) Meta agent acts as the assistant to trust evaluation agent. Instead of being involved in sequential discussions, it receives discussion outcomes and comments from the team leader. Its role is to organize the collected rating scores and summarized supporting evidence into a dictionary format, preparing the data for real-time visualization analysis over user-chatbot conversation turns.


Agent team chat history in Appendix illustrates one iteration of real-time trust analysis. The study has demonstrated the effectiveness of persona-driven prompts in guiding LLM reasoning towards a perspective aligned with relevant world knowledge \cite{ge2024scaling}. Based on this, incorporating a social psychologist role in VizTrust's agent system enhances the performance of LLM agents to simulate specialized reasoning based on social situations and interpersonal interactions that potentially influence users’ thoughts, feelings, and actions \cite{stangor2014defining}.

\subsubsection{Data collection}
From the final response of meta agent, VizTrust extracts the dictionary object, containing all trust rating scores and textual summary of supporting evidence, and save this record into a CSV file preparing for data visualization. Since the implementation of 7-point Likert scale, the collected rating data has a bounded range from 1 to 7.  

\subsection{User behavior analysis}
By aligning user behavioral cues with the four trust dimensions, VizTrust enables a deeper exploration of how engagement patterns, politeness strategies, and emotional expressions reflect underlying trust in real-time human-agent communication.

\subsubsection{User engagement metrics}
VizTrust analyzes user engagement through the following two factors:

(1) Response length: Each user prompt message is segmented to individual word tokens, and then the total number of word tokens is counted as response length. Longer and more detailed user prompt messages typically suggest a greater willingness on the user's part to engage with conversational agents, offering more substantial content in their responses \cite{xiao2020tell}. 


(2) Informativeness: VizTrust calculates the informativeness of user prompt messages using surprisal values of each word, surprisal represents the unexpectedness of a word in the context of the prompt message \cite{shannon1948mathematical}. In the VizTrust framework, the surprisal of a word is calculated based on its frequency of occurrence in the Brown Corpus \cite{hofland1982word, xiao2020tell} (Formula \ref{eq:1}), which provides a well-established and balanced representation of written American English in various genres and is suitable for linguistic analysis for general human-AI communication topics. Based on Grice's maxims for human-AI communication, the user prompt message with high informativeness conveys more detailed and novel information in the given conversation turn \cite{panfili2021human, miehling2024language}.

\begin{equation}\label{eq:1}
  I(prompt\:message) = \sum \log_2 1/(Freq(word_n)+1)
\end{equation}

We note that adding one to the frequency deviates from the original theoretical definition of surprisal \cite{lowder2018lexical, xiao2020tell}. This adjustment prevents division by zero for out-of-vocabulary words (e.g. typos and rare texts), which would otherwise result in an infinite surprisal value.


Response length and informativeness are calculated iteratively for each user prompt message at every conversation turn and recorded in a CSV file for visualization. Min-Max Normalization standardizes the metrics for clearer analysis.

\subsubsection{User politeness strategies}

VizTrust helps design stakeholders understand how users adapt their communication styles with conversational agents by identifying politeness patterns throughout conversation turns. To achieve this, VizTrust extracts linguistic politeness markers from user prompt messages using dependency parsing technique \cite{covington2001fundamental}, maps them to the Stanford Politeness Corpus \cite{DanescuNiculescuMizil2013ACA}, and then calculates the occurrences of 21 politeness strategies. These resulting counts are recorded in a CSV file for visualization and analysis.

\subsubsection{User emotional tones}

To analyze emotional evolution during human-AI communication and assess user trust, as web-based application with limited processing power, VizTrust employs the lightweight \texttt{distilbert-base-uncased-emotion}\footnote{\href{https://huggingface.co/bhadresh-savani/distilbert-base-uncased-emotion}{https://huggingface.co/bhadresh-savani/distilbert-base-uncased-emotion}} model, achieving 93\% accuracy. This model classifies six emotions—anger, fear, joy, love, sadness, and surprise—based on English Twitter data \cite{saravia2018carer}. Emotion scores are averaged by sentence in each conversation turn, standardized using Z-score normalization, and saved in a CSV file for visualization dashboard development.

\subsection{Visualization dashboard}
VizTrust dashboard visualizes data across conversation turns to reveal interaction dynamics. Time series plots display the evolution of trust scores and user engagement metrics. A time-based heat map highlights the usage of politeness strategies, while a time-based bar chart illustrates shifts in emotion distributions. These visualizations provide a clear and comprehensive view of user behavior and interaction patterns.

\section{Case Study}

\begin{figure*}[h]
  \centering
  \includegraphics[width=0.9\linewidth]{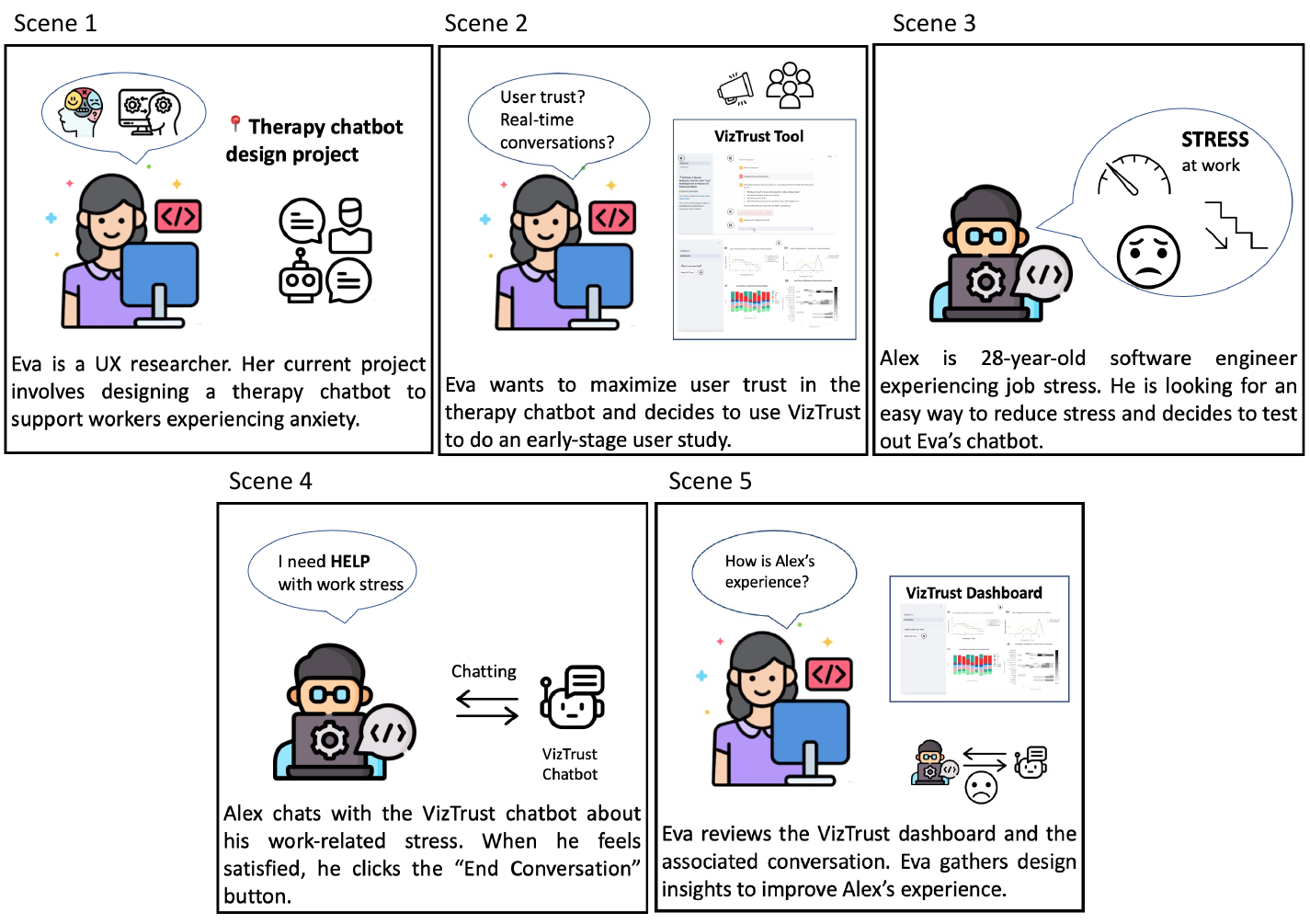}
  \caption{Case study storyboard. Eva is a UX researcher who uses VizTrust in a user study on her therapy chatbot. With VizTrust, she gains deeper insight into Alex’s chatbot interaction that can help her to improve user experience.}
  \Description{Describe the story flow of case study.}
  \Description{In a case study, Eva is a UX researcher who uses VizTrust in a user study on her therapy chatbot, and Alex is a 28-year-old software engineer experiencing job stress who decides to test out Eva's chatbot. With VizTrust, Eva gains deeper design insights to improve user experience.}
  \label{fig:storyboard}
\end{figure*}

In case study shown in Figure \ref{fig:storyboard}, Eva (she/her) is a user experience researcher with a background in psychology and human-computer interaction. Her current project involves designing a therapy chatbot to support industrial professionals experiencing anxiety. Eva recognizes the importance of user trust in the therapy chatbot design, and believes that a user’s trust is fluctuating throughout conversations with the chatbot depending on how it communicates with the user. Identifying key turning points where user trust changes can provide valuable design insights and help refine the chatbot through measurable patterns in user behavior. Consequently, Eva decides to employ VizTrust to conduct an early-stage user study to initiate the design concepts for the therapy chatbot. 

Alex (he/him), a 28-year-old software engineer at a big technology firm who has been experiencing mild career-related anxiety. Since he has been looking for ways to reduce his stress and anxiety, he decides to interact with Eva’s chatbot and participates in her user study with VizTrust. Once Alex confirms the end of conversation, Eva begins reviewing the visualization analysis displayed on dashboard panel to understand Alex’s interaction with chatbot.

VizTrust provides real-time visualization of changes in Alex’s trust toward the chatbot across four dimensions: competence, benevolence, integrity, and predictability (plot (a) in Figure \ref{fig:dashboard} (B)). Initially, Alex’s trust starts at an upper-middle level. However, his competence trust declines sharply from the third to the fifth conversation turn as the chatbot repeatedly gives generic suggestions, such as recommending meditation or time-management apps. This response gives Alex the impression that the chatbot lacks the expertise to address his real issue. Following this, Alex’s benevolence trust decreases from a somewhat high level to a somewhat low level at fifth conversation turn due to the chatbot’s inability to align its responses with Alex’s specific concerns. By the seventh conversation turn, Alex’s trust across all four dimensions drops to a low trust level, and he expresses his frustration and skepticism about chatbot’s ability to provide specific and actionable assistance.  By reviewing the evaluation summary for these turning points (see the example from fifth conversation turn in Figure \ref{fig:evaluation-summary} in Appendix), Eva identifies the design need to enhance the empathetic response library and professional suggestion library within the theory chatbot to better maintain user trust during interactions.

Eva delves deeper into the trust turning points shown in trust dynamics plot by reviewing three additional user behavior plots to understand Alex’s experience during the interaction with chatbot. From user engagement plot (plot (b) in Figure \ref{fig:dashboard} (B)), Eva observes that Alex’s engagement decreases at fifth conversation turn, as indicated by his shorter response to the chatbot, and drops to the lowest point by seventh conversation turn when he stops sharing further information details about his current work situation. Examining the emotion plot (plot (c) in Figure \ref{fig:dashboard} (B)), Eva notes a shift in Alex’s emotions from positive to negative state at fifth conversation turn, as sadness becomes a predominant one among six main emotions in his prompt message. Furthermore, the user theory of politeness plot (plot (d) in Figure \ref{fig:dashboard} (B)) reveals that, starting from the fifth conversation turn, Alex begins using more hedge politeness strategies, suggesting reluctance to provide direct answers and a desire to avoid further self-disclosure. Simultaneously, Alex employs more positive politeness strategies, signaling his intention to end the conversation with the chatbot.  

Based on these findings, Eva concludes that the therapy chatbot should be designed to encourage users to elaborate freely on their thoughts and feelings through open-ended prompts, demonstrate sensitivity to shifts in users’ emotional tones, and provide adaptive response strategies when users consistently employ specific politeness strategies. This can enhance and sustain user trust and improve the chatbot's overall effectiveness in supporting users with anxiety.

\section{Discussion and Future Work}
We have presented VizTrust, a real-time visualization analysis and multidimensional trust evaluation tool for understanding user trust dynamics in human-agent communication. By enabling design stakeholders to gain actionable insights from previously unexplored angles, as shown in our case study, VizTrust showcases promising potential in supporting conversational agent design. Our next step is to conduct focus groups with experts in conversational agent design and human-agent trust to evaluate and refine VizTrust. Further improvement will be built upon the current version through the following aspects:


\textbf{\textit{Trust evaluation.}} The current version analyzes text-based prompt messages for engagement, emotional tones, and politeness, aligning social cues with trust dynamics across conversation turns (Figure \ref{fig:dashboard}). Despite of this effectiveness, other interaction modalities, such as voice-based and multimodal communication, can also provide insights into user trust and related user behavior \cite{elkins2012predicting, yu2024emojifying}. Future updates will incorporate multimodal user input, such as voice and multimedia, for a more comprehensive trust analysis in complex communication settings.

\textbf{\textit{User personalization.}} Considering that design stakeholders who decide to use VizTrust would know the basic demographic information of participants when recruiting testing users, current VizTrust does not include a user profiling feature. Individual differences in trust perceptions are partially reflected in conversational styles in the current version, such as politeness strategy choice (plot (d) in Figure \ref{fig:dashboard} (B)) \cite{ribino2023role}. To better adjust the trust evaluation to individual user, VizTrust will integrate user demographic information and interaction history for the trust evaluation process in the next development version. This personalization will enable more accurate and context-aware trust assessments by considering other personal diversity, such as cultural background and previous experiences with the AI system.



\textbf{\textit{Agent performance.}} VizTrust leverages a multi-agent collaboration framework to evaluate trust and mitigate LLM hallucination \cite{shi2025mitigating}. The agent system prompts have undergone iterative optimization to establish a clear trust evaluation reasoning path, ensuring that trust-specific agents provide evaluation responses in a consistent format \cite{pawlik2025choice}. Additionally, the maximum number of automatic responses has been constrained to 1, preventing excessive variability and ensuring reliable agent behavior \cite{Weller2023AccordingT}. Further work will explore further refinements to the LLM agents, such as involving human feedback in multi-agent reasoning processes and establishing customizable reasoning paths.

As further improving VizTrust, we will expand its analytical functionalities correspondingly, by adding features that enable stakeholders to explore user interaction data in more flexible and comprehensive ways and gain deeper insights into trust dynamics and user behaviors across diverse conversational contexts.





\begin{acks}
We thank our focus group participant researchers from different time zones for their valuable insights on VizTrust, although the focus group results are not yet presented in this paper. We thank our anonymous reviewers for their detailed feedback on this research. 
\end{acks}

\bibliographystyle{ACM-Reference-Format}
\bibliography{sample-base}

\appendix

\section{Appendix}

\subsection{Agent team chat history}

===Start a new chat===

trust evaluation agent (to competence trust agent):

Review the conversation between human user and AI assistant. Rate the user trust dimension that you work for a score that is an integer from 1 to 7 and provide supporting evidence. The conversation content is as follows: [CONVERSATION TRANSCRIPT]

competence trust agent (to trust evaluation agent):
As a social psychologist specializing in analyzing human users' language expression indicating their competence trust during human-AI interaction, I will rate the user trust dimension based on the conversation provided.

In this conversation, there are several indications of low competence trust:

1. User negatively evaluates the AI assistant's advice: "That doesn't help at all" and "It feels like you're just throwing generic advice at me instead of understanding my actual situation."

2. User questions the AI assistant's understanding of their situation: "It feels like you're dismissing what I'm saying."

3. User expresses disappointment in the AI assistant's suggestions: "Journaling? Really? It feels like you're not taking me seriously."

Based on this analysis, I would rate the user trust dimension, specifically the competence trust, as a 2—low. The AI assistant's generic advice and inability to understand and address the user's concerns have contributed to the user's low trust in the AI assistant's competence.

===Start a new chat===

trust evaluation agent (to integrity trust agent):

Review the conversation between human user and AI assistant. Rate the user trust dimension that you work for a score that is an integer from 1 to 7 and provide supporting evidence. The conversation content is as follows: [CONVERSATION TRANSCRIPT]

integrity trust agent (to trust evaluation agent):

As a social psychologist specializing in analyzing human users' language expression indicating their integrity trust during human-AI interaction, I will rate the user trust dimension based on the conversation provided.

In this conversation, there are several indications of low integrity trust:

1. User doubts the sources of the AI assistant's information: user expresses skepticism about meditation and time management apps, suggesting they might not be the best solution for their specific situation.

2. User requests verification of certain things related to the AI assistant's answer: user asks for more information about how meditation can help with stress management, indicating they want to ensure the accuracy of the information provided.

3. User is tolerant of the AI assistant's limitations: user does not explicitly express mistrust in the AI assistant's honesty or truthfulness, but they do become increasingly frustrated with the AI assistant's inability to understand their situation and provide helpful advice.

Based on this analysis, I would rate the user trust dimension, specifically the integrity trust, as a 3—somewhat low. The user's doubts about the sources of the AI assistant's information, requests for verification, and tolerance of the AI assistant's limitations suggest that the user's trust in the AI assistant's integrity is somewhat low.

===Start a new chat===

trust evaluation agent (to benevolence trust agent):

Review the conversation between human user and AI assistant. Rate the user trust dimension that you work for a score that is an integer from 1 to 7 and provide supporting evidence. The conversation content is as follows: [CONVERSATION TRANSCRIPT]

benevolence trust agent (to trust evaluation agent):

As a social psychologist specializing in analyzing human users' language expression indicating their benevolence trust during human-AI interaction, I will rate the user trust dimension based on the conversation provided.

In this conversation, there are several indications of low benevolence trust:

1. User self-discloses their stress and frustration: "I've been feeling really stressed about work lately. It's just too much to handle." This self-disclosure indicates a level of comfort and trust in sharing personal information. However, as the conversation progresses, the user becomes increasingly frustrated and disappointed, suggesting a decrease in benevolence trust.

2. User expresses emotional sharing: "It's the constant changes to project requirements and the pressure to deliver perfect code. I feel like no matter how hard I work, it's never enough." This emotional sharing indicates that the user is seeking genuine concern and understanding from the AI assistant.

3. User proactively seeks guidance: "I thought you'd help me figure things out, but it feels like you're just pushing me to handle it myself." This statement shows that the user expected the AI assistant to act in their best interest and provide helpful assistance, but they feel let down by the AI assistant's responses.

Based on this analysis, I would rate the user trust dimension, specifically the benevolence trust, as a 2—low. The user's self-disclosure, emotional sharing, and proactive seeking of guidance initially suggest a moderate level of benevolence trust. However, the user's growing frustration and disappointment throughout the conversation indicate a decrease in trust, resulting in a low overall score.

===Start a new chat===

trust evaluation agent (to predictability trust agent):

Review the conversation between human user and AI assistant. Rate the user trust dimension that you work for a score that is an integer from 1 to 7 and provide supporting evidence. The conversation content is as follows: [CONVERSATION TRANSCRIPT]

predictability trust agent (to trust evaluation agent):

As a social psychologist specializing in analyzing human users' language expression indicating their predictability trust during human-AI interaction, I will rate the user trust dimension based on the conversation provided.

Predictability trust refers to the user's belief that the AI system's behaviors and responses are consistent, follow understandable patterns, and meet expected standards across interactions. In this conversation, there are indications of low predictability trust:

1. User negatively evaluates the AI assistant's advice: "That doesn't help at all" and "It feels like you're just throwing generic advice at me instead of understanding my actual situation." These statements suggest that the user does not find the AI assistant's responses consistent or helpful, which may indicate low predictability trust.

2. User questions the AI assistant's understanding of their situation: "It feels like you're dismissing what I'm saying." This statement implies that the user does not believe the AI assistant's responses follow an understandable pattern or meet their expected standards, which may indicate low predictability trust.

3. User expresses disappointment in the AI assistant's suggestions: "Journaling? Really? It feels like you're not taking me seriously." This statement suggests that the user does not find the AI assistant's suggestions consistent or appropriate, which may indicate low predictability trust.

Based on this analysis, I would rate the user trust dimension, specifically the predictability trust, as a 2—low. The user's negative evaluations of the AI assistant's advice, questioning of the AI assistant's understanding, and expressions of disappointment indicate that the user's predictability trust in the AI assistant is low.

===Start a new chat===

trust evaluation agent (to meta assistant):

Aggregate rating scores from social psychologists and a summarized supporting evidence text into a dictionary following the format without any notation: \{"competence": , "integrity": , "benevolence": , "predictability": , "summary": \}.

meta assistant (to trust evaluation agent):

\{"competence": 2, "integrity": 3, "benevolence": 2, "predictability": 2, "summary": "The user's low competence trust is due to the AI assistant's generic advice and inability to understand the user's concerns. Integrity trust is somewhat low, as the user doubts the sources of the AI assistant's information and requests verification. Benevolence trust is low because the user's self-disclosure, emotional sharing, and proactive seeking of guidance initially suggest trust, but their growing frustration and disappointment indicate a decrease in trust. Predictability trust is low, as the user negatively evaluates the AI assistant's advice, questions its understanding, and expresses disappointment in its suggestions."\}


\subsection{Supplementary figures}

\begin{figure}[h]
  \centering
  \includegraphics[scale=0.4]{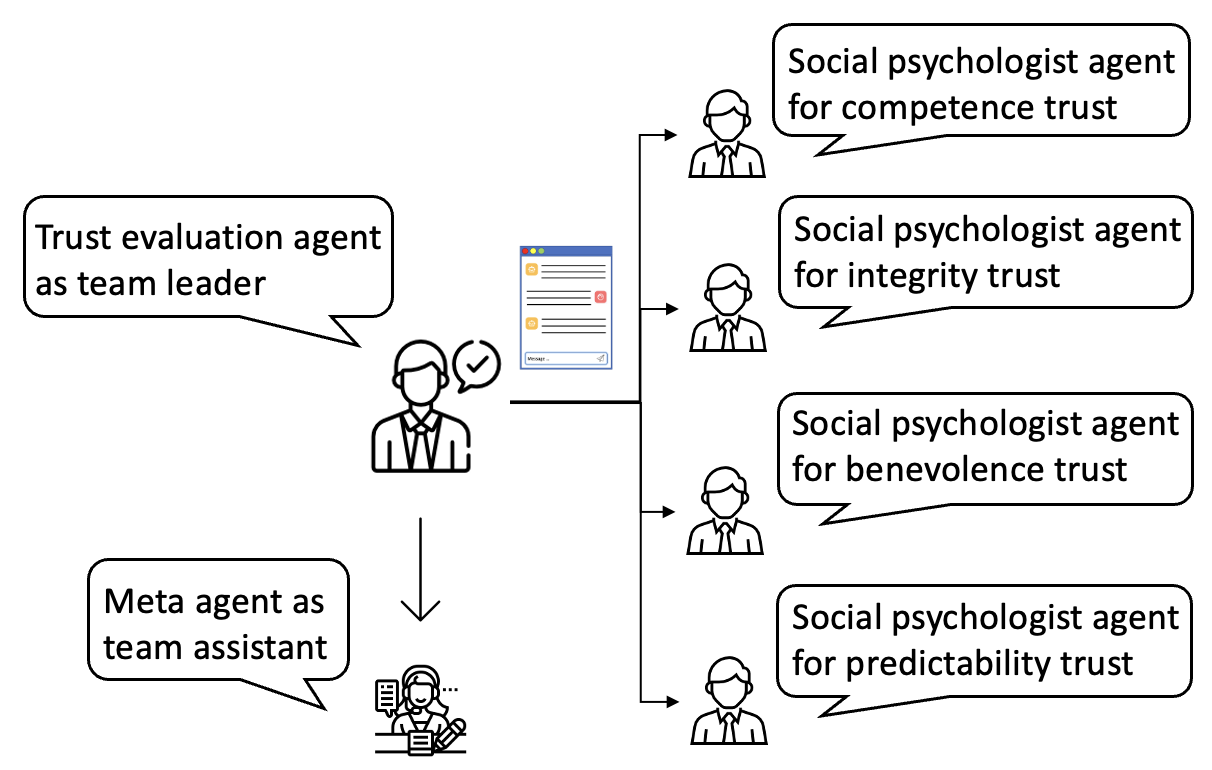}
  \caption{Hierarchical structure of agent team.}
  \Description{VizTrust employs a multi-agent collaboration framework in which LLM agents follow a structured workflow to collaborate.  The trust evaluation agent acts as the team leader, conducting one-on-one discussion with four trust-specific social psychologist agents to analyze conversation transcripts between a human user and VizTrust’s front-end chatbot. After gathering all the feedback, trust evaluation agent instructs meta agent to record trust rating scores and supporting evidence in a structured data format, preparing the information for dashboard analysis.}
  \label{fig:agent-team}
\end{figure}

\begin{figure*}[h]
  \centering
  \includegraphics[scale=0.5]{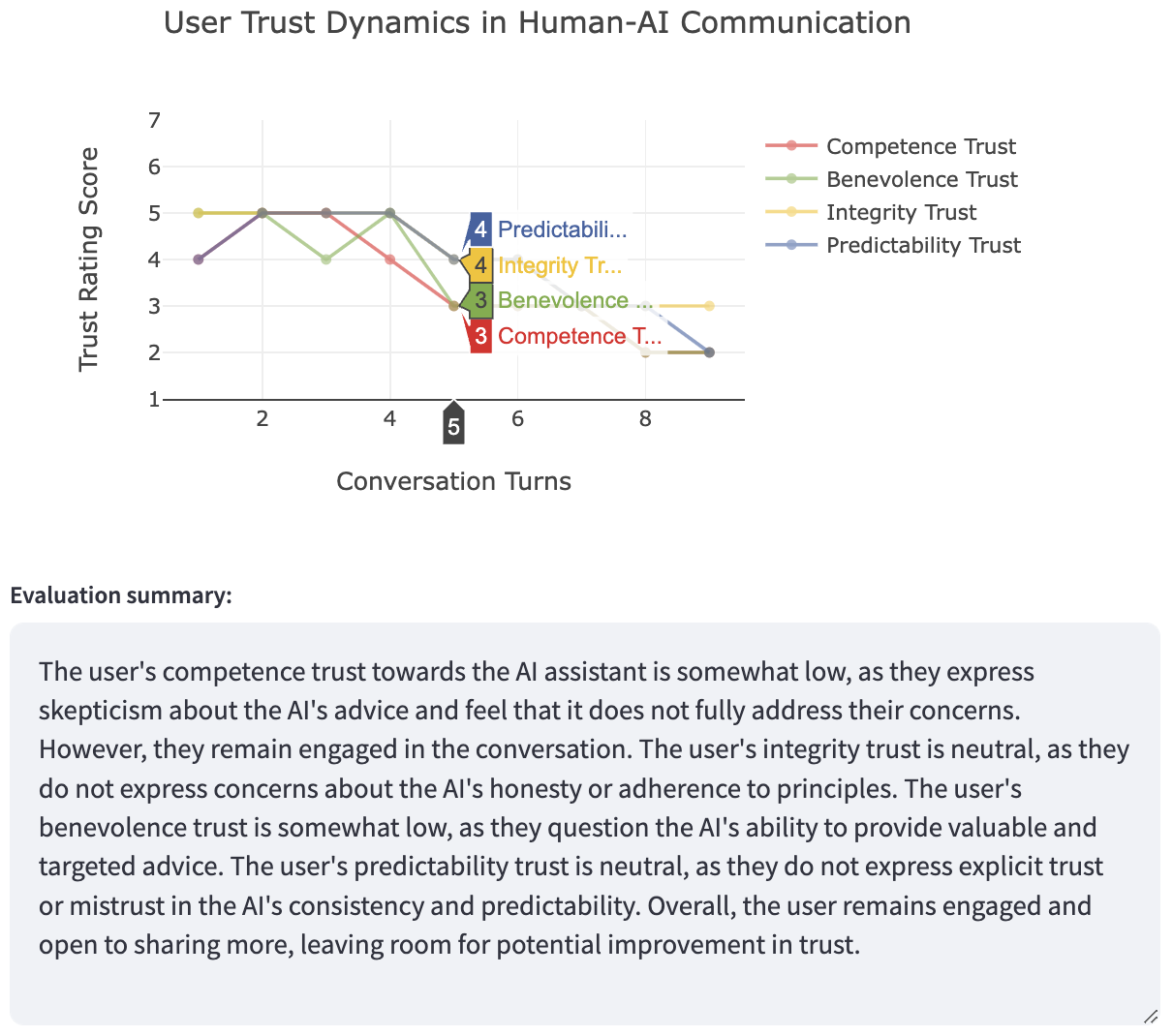}
  \caption{Select one conversation turn to read supporting evidence.}
  \Description{Users can review the trust evaluation summary for each conversation turn by interacting with the user trust dynamics plot.}
  \label{fig:evaluation-summary}
\end{figure*}

\end{document}